\documentclass[aps,prl,twocolumn,groupedaddress,showkeys,showpacs]{revtex4}

\usepackage{amssymb,amsbsy,amsmath}
\usepackage{multirow}
\usepackage{graphicx}

\newcommand{\op}[1]{%
    \fontdimen12\textfont3=2pt\fontdimen12\scriptfont3=1.4pt%
    \!\null\mathop{\vphantom{#1}\smash{#1}}\limits_{\sim}\null\!}
\newcommand{\xref}[1]{\protect\ref{#1}}
\newcommand{\figref}[1]{Fig.~\protect\ref{#1}}
\newcommand{\fmref}[1]{(\protect\ref{#1})}

\def\ket#1{\, | \, {#1} \, \rangle}

\newcommand {\mofe} {\{$\textrm{Mo}_{72}\textrm{Fe}_{30}$\}}
\newcommand {\mocr} {\{$\textrm{Mo}_{72}\textrm{Cr}_{30}$\}}
\newcommand {\mov} {\{$\textrm{Mo}_{72}\textrm{V}_{30}$\}}

\begin{document}

\title{Optimized perturbation theory for molecular antiferromagnets}

\author{Roman Schnalle}
\email{rschnall@uos.de}
\affiliation{Universit\"at Osnabr\"uck, Fachbereich Physik,
D-49069 Osnabr\"uck, Germany}

\author{J\"urgen Schnack}
\email{jschnack@uni-bielefeld.de}
\affiliation{Universit\"at Bielefeld, Fakult\"at f\"ur Physik,
  Postfach 100131, D-33501 Bielefeld, Germany}

\date{\today}

\begin{abstract}
The energy spectra of mesoscopic, i.e. few-body quantum systems
are of great interest in several areas of physics such as
nuclear physics, cluster physics or magnetism. One way to obtain
an approximate spectrum is to diagonalize with reduced but
carefully chosen basis sets. In this Letter we propose for the
case of mesoscopic antiferromagnets to use the eigenstates of
the rotational band  Hamiltonian for this purpose. These states
are not only well adapted, they can also be easily constructed. 
\end{abstract}

\maketitle

\emph{Introduction}---The spectrum of many interesting molecular
antiferromagnets is theoretically often inaccessible due to the
prohibitive size of the underlying Hilbert space. Experimentally
the system of interest might very well be accessible for
instance by neutron scattering, EPR, specific heat or
magnetization measurements. The interpretation of the
experimental data thus suffers from numerical restrictions.  A
prominent example for this problem is given by the giant
Keplerate molecules \mofe\ \cite{MSS:ACIE99}, \mocr\
\cite{TMB:ACIE07}, and \mov\ \cite{MTS:AC05}, in which 30
paramagnetic ions occupy the vertices of an icosidodecahedron
and interact by nearest-neighbor antiferromagnetic
exchange. These molecules are rather similar to the Kagome
lattice antiferromagnet. Thus by studying them one can also gain
further insight into the properties of the Kagome lattice
antiferromagnet.

In this Letter we propose to apply an approximate
diagonalization that is guided by perturbation theory
arguments. It rests on the observation that the low-lying
spectrum of many finite size, for instance molecular,
antiferromagnets can be rather successfully approximated by
so-called rotational bands (or towers of states)
\cite{ScL:PRB00,SLM:EPL01,Wal:PRB01,BLL:PRB94,GSS:PRB89}, which
are the eigenstates of the rotational band Hamiltonian. We
suggest to use these basis states for an approximate
diagonalization of the full Heisenberg Hamiltonian. We argue
that such a procedure should be rather good since in a
perturbation theory picture the rotational band Hamiltonian
plays the role of the unperturbed Hamiltonian which constitutes
already the major part of the Hamiltonian. Thus one can consider
the eigenstates of the rotational-band Hamiltonian as
prediagonalized basis states for which the full Hamiltonian
should have not too big off-diagonal matrix elements
anymore. Similar ideas are currently applied in nuclear physics,
see for instance Ref.~\cite{RoN:PRL07}. The aim of our attempt
is to obtain sufficiently many approximate eigenstates in order
to accurately estimate thermodynamic functions depending on
temperature and applied magnetic field. This constitutes a major
difference to other approximate methods such as Density Matrix
Renormalization Group techniques (DMRG)
\cite{Whi:PRB93,Sch:RMP05}, that yield only a few states and
practically are restricted to one-dimensional systems, and to
Lanczos \cite{Lan:JRNBS50} or finite temperature Lanczos
\cite{ZST:PRB06} methods, that currently cannot be applied in
Hilbert spaces with dimensions bigger than approximately
$10^9$. Our proposed method is also not restricted by the
negative sign problem in contrast to Quantum Monte Carlo methods
\cite{HeS:PRB00}. 

As further advantages of the proposed method we would like to
mention that the approximate basis states can be constructed
according to spin coupling schemes and that the matrix elements
of the Hamiltonian matrix can then be evaluated using
Irreducible Tensor Operator (ITO) techniques
\cite{GaP:GCI93,BCC:IC99,Wal:PRB00}. In addition point group
symmetries can be applied, which altogether provides a much
better characterization of the approximated spectrum than other
approximate methods can deliver.  We demonstrate the
applicability of our approximate diagonalization as well as
limitations by several examples.

\emph{Theoretical Method}---It turns out that many magnetic
molecules can be well described within the Heisenberg model,  
\begin{eqnarray}
\label{E-2-1}
\op{H}
&=&
-
\sum_{u, v}\;
J_{uv}\,
\op{\vec{s}}(u) \cdot \op{\vec{s}}(v)
\ .
\end{eqnarray}
Here the sum reflects the exchange interaction between
spins given by spin operators
$\op{\vec{s}}(u)$ at sites
$u$. A negative value of $J_{uv}$
corresponds to antiferromagnetic coupling. For the following
discussions an antiferromagnetic
nearest-neighbor exchange of constant size $J$ is assumed.

Since the Hamiltonian commutes with the total spin, we can find
a common eigenbasis $\{\ket{\nu} \}$ of $\op{H}$, $\op{S}^2$,
and $\op{S}_z$ and denote the related eigenvalues by $E_{\nu}$,
$S_{\nu}$, and $M_{\nu}$, respectively. For not too large spin
systems all eigenvalues can be obtained numerically, but with
growing system size this method becomes impossible since the
related Hilbert space grows as $(2s+1)^N$ for $N$ spins of spin
quantum number $s$.

\begin{figure}[ht!]
\centering
\includegraphics[clip,width=65mm]{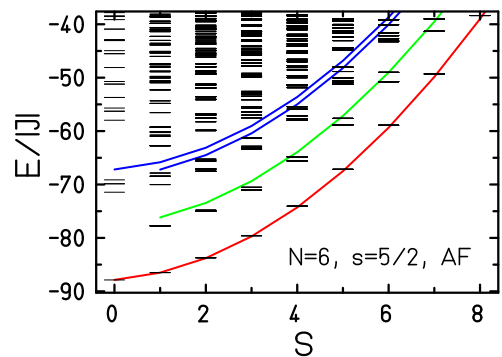}
\caption{(Color online) Low-lying spectrum of an
  antiferromagnetic ring with $N=6$ and $s=5/2$: The exact
  energy eigenvalues of \eqref{E-2-1} are depicted by dashes. The
  rotational band energies according to \eqref{E-2-2} are given
  by curves in order to highlight their parabolic shape which is
  typical for a rotor.}
\label{F-1}
\end{figure}

Nevertheless, it has been realized that the low-energy spectrum
of many antiferromagnetically coupled spin systems can be rather
well approximated by so-called rotational bands, which resemble
the spectrum of a quantum rotor \footnote{Also in this respect
there is a clear analogy to nuclear physics where such bands are
termed Yrast bands.}. This is especially true for
antiferromagnets that consist of sublattices, i.e. where spins
on one sublattice interact only with those of other sublattices.
Prominent examples are given by ring molecules with an even
number of spins that are bipartite (two sublattices) or by
clusters of cuboctahedral or icosidodecahedral structure that
possess three sublattices.  In such cases the Heisenberg
Hamiltonian can be approximated by
\begin{eqnarray}
\label{E-2-2}
\op{H}^{\text{rb}}
&=&
-\frac{D\, J}{N}\,
\left[
\op{\vec{S}}^2 - 
\sum_{n=1}^{N_s} \op{\vec{S}}_n^2
\right]
\ ,
\end{eqnarray}
where the operators $\op{\vec{S}}_n$ are the total spins of the
sublattices of the spin system. The prefactor $DJ/N$ is the
effective coupling strength of the sublattice spins
\cite{ScL:PRB00}.  The eigenvalues of this Hamiltonian are
easily evaluated since total spin and sublattice spins mutually
commute. The multiplicity of the energy levels is given by the
various ways to couple the individual spins to the
sublattice spins and the sublattice spins to the total spin.  It
has been shown that such a rotational band Hamiltonian describes
the low-lying levels of bipartite systems with
very high accuracy \cite{ScL:PRB00,SLM:EPL01,Wal:PRB01}, even
neutron scattering data can be analyzed without diagonalizing
the full Hamiltonian \cite{WGC:PRL03}. In the case of frustrated
three- and four-sublattice \cite{HeZ:PRL98,Hen:PRL06} systems
such an assumption -- although less accurate -- can still
explain gross properties like the magnetization \cite{SLM:EPL01}
or the position of resonances in inelastic neutron scattering
\cite{GNZ:PRB06}. Also in the context of finite square and
triangular lattices of spin-$1/2$ Heisenberg antiferromagnets
such an approximation has been verified
\cite{BLL:PRB94,GSS:PRB89}. Figure \xref{F-1} shows as an
example the low-lying exact spectrum as well as the rotational
band energies for an antiferromagnetic ring with $N=6$ and
$s=5/2$.

The idea we propose in this Letter is to use the eigenstates of
the rotational band Hamiltonian \eqref{E-2-2} for a
diagonalization of the full Hamiltonian
\eqref{E-2-1}. Technically this is done with Irreducible Tensor
Operator (ITO) techniques \cite{GaP:GCI93,BCC:IC99}. If one
could take all eigenstates this would of course correspond to an
exact diagonalization. But in cases where this is not possible,
a subset of low-lying eigenstates of $\op{H}^{\text{rb}}$ can
still be used for an approximate diagonalization. The
expectation is that such a procedure yields a largely improved
energy spectrum compared to the plain rotational band spectrum
or to spectra obtained by DMRG or Lanczos. The reason for that
is that from a perturbation theory point of view
\begin{eqnarray}
\label{E-2-3}
\op{H}
&=&
\op{H}^{\text{rb}}
+
\op{H}^{\prime}
\ ,
\end{eqnarray}
and $\op{H}^{\prime}$ is relatively small compared to
$\op{H}^{\text{rb}}$. Therefore, the rotational band model works
as a guide to select a finite set of basis states for an
approximate diagonalization. The fact that the eigenstates of
$\op{H}^{\text{rb}}$ do already possess spin rotational symmetry
further improves the approximation.

With the help of examples we will demonstrate in the next part
how good such an approximate diagonalization is, but some
general remarks can be made already at this point. It was
already observed that the rotational band approximation is much
better for bipartite antiferromagnets, since they are not
frustrated, as for systems with three or more sublattices. One
can expect that also the approximate diagonalization using
rotational band states will follow the same trend. The
approximate diagonalization yields the new eigenstates as linear
superposition of rotational band states. The influence of
higher-lying rotational band levels on low-lying eigenstates
should decrease with their energy difference as in perturbation
theory where the energy difference appears in the
denominator. Unfortunately, this can in principle be (over-)
compensated if the density of states grows strongly with
increasing energy and if the Hamiltonian still connects states
even if they are far apart in energy. In the positive spirit of
perturbation theory we hope that not both "ifs" have to be
answered with yes. But even if in certain cases the absolute
energy eigenvalue is only poorly met it could nevertheless be
that the relative energies of the approximate levels are much
better, which would be sufficient for a thermodynamic
discussion.

\emph{Numerical examples}---As a first example we discuss
antiferromagnetic spin rings. For an even number of sites one
expects rather good results since these systems do not suffer
from frustration. In order to quantify the quality of the
method we investigate the convergence of the energy levels as a
function of the number of used rotational band states. For this
purpose we group these states according to their energy band,
i.e. according to the quantum numbers of their total and
sublattice spins. 

\begin{figure}[ht!]
\centering
\includegraphics[clip,width=65mm]{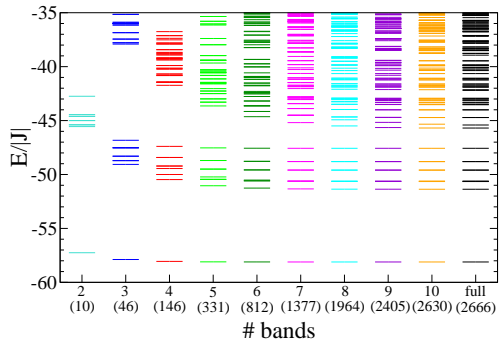}
\caption{(Color online) Convergence of the approximate energy
  levels as a function of the number of used rotational bands
  for a spin ring of $N=8$ spins $s=5/2$.}
\label{F-2}
\end{figure}

Figure \xref{F-2} shows how the levels of the subspace $S=0$
converge against the complete exact spectrum for a spin ring of
$N=8$ spins $s=5/2$. The numbers given at the $x$-axis denote
the number of rotational bands used for the approximate
diagonalization as well as in parenthesis the number of states
contained in these bands. One realizes that the convergence is
fast, and that the exact energy spectrum is rather well
approximated using only 4 to 5 rotational bands for the
diagonalization.

\begin{figure}[ht!]
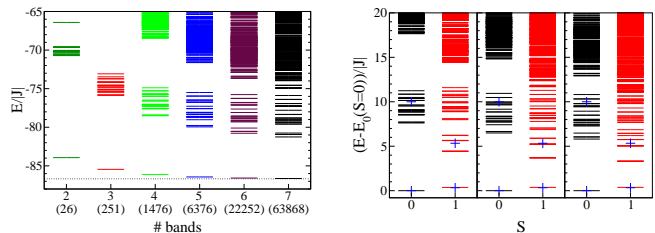

\centering
\includegraphics[clip,width=40mm]{fig-3a.eps}
$\quad$
\includegraphics[clip,width=40mm]{fig-3b.eps}
\caption{(Color online) L.h.s.: Convergence of the approximate
  energy levels as a function of the number of used rotational
  bands for a spin ring of $N=12$ spins $s=5/2$.
  R.h.s. Approximate energy levels in subspaces $S=0$ and $S=1$
  for 4, 5, and 6 used rotational bands. The crosses denote the
  positions of the two lowest rotational band energies for $S=0$
  and $S=1$ ($D=4$ in \fmref{E-2-2}).} 
\label{F-3}
\end{figure}

As an example for a large system we show the approximate
spectrum for a ring of $N=12$ spins $s=5/2$ in
\figref{F-3}. Such a system was synthesized an the basis of
Fe$^{\text{III}}$ ions \cite{CCF:ACIE99}. Its parameters were
determined using the rotational band approximation
\cite{AII:PB03,IAA:JPSJ03} and later improved by means of QMC
\cite{EnL:PRB06}. On the l.h.s. of \figref{F-3} we show how the
approximate energies converge in the subspace of $S=0$ which has
a dimension of 1,949,156. The r.h.s. of \figref{F-3} displays
the subspaces of $S=0$ and $S=1$ together in order to
demonstrate how the low-lying gap structure develops with the
size of the truncated basis. We find that using about six
bands, i.e. about 60,000 basis states provides a reasonable
low-energy spectrum.

The understanding of the low-lying spectrum of the giant
Keplerate molecules \mofe\ \cite{MSS:ACIE99}, \mocr\
\cite{TMB:ACIE07}, and \mov\ \cite{MTS:AC05}, in which 30
paramagnetic ions with spins $s=5/2, 3/2, 1/2$, respectively,
occupy the vertices of an icosidodecahedron, is of great
importance due to its similarity with the kagome lattice
antiferromagnet, see \cite{RLM:08} for a recent review. For
example, an interpretation of low-energy inelastic neutron
scattering (INS) data in terms of rotational bands
\cite{GNZ:PRB06} or spinwave theory \cite{CeZ:PTPS05} has been
only partially successful. A first improvement could be obtained
in \cite{WAL:PRB07} where an approximate diagonalization using
the states of just the two lowest rotational bands was
performed. Nevertheless, the large low-lying density of states
visible in the INS data was not reflected by any of the
approximate spectra. Also for the interpretation of $\mu$SR data
for \mofe\ the rotational band model provides a first but not
yet satisfactory approximation \cite{LMC:PRB07}.

\begin{figure}[ht!]
\centering
\includegraphics[clip,width=65mm]{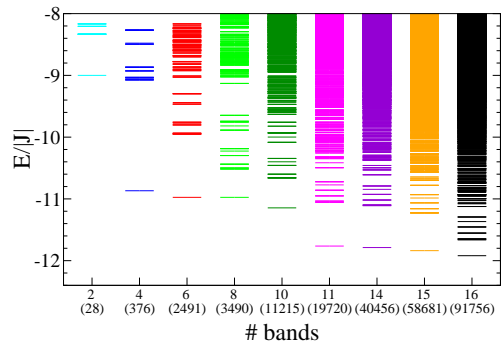}
\caption{(Color online) Convergence of the approximate energy
  levels as a function of the number of used rotational bands
  for an icosidodecahedron ($N=30$) of spins $s=1/2$.}
\label{F-4}
\end{figure}

Figure \xref{F-4} demonstrates that in this respect the proposed
approximate diagonalization can yield a substantial
improvement. The figure shows how the energy spectrum in the
sector of total spin $S=0$ converges with the number of used
rotational band states. Since this structure belongs to the
geometrically most frustrated ones -- Icosidodecahedron, Kagome,
Pyrochlore -- it is no surprise that the convergence is much
slower than in the case of bipartite spin systems. In addition
the convergence is also rather irregular, especially for the gap
between the ground state and the first excited one in this
subspace. But with nowadays numerical diagonalization
capabilities one is able to arrive at a level of approximation
that explains the high density of low-lying states, that is
believed to be present in such highly frustrated
antiferromagnets \cite{AHL:PRL98}.

\begin{figure}[ht!]
\centering
\includegraphics[clip,width=65mm]{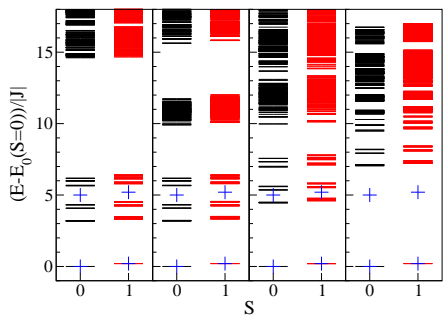}
\caption{(Color online) Approximate energy levels in subspaces
  $S=0$ and $S=1$ for 8, 9, 10, and 11 used rotational bands for an
  icosidodecahedron ($N=30$) of spins $s=5/2$. The crosses
  denote the positions of the two lowest rotational band
  energies for $S=0$ and $S=1$ ($D=6$ in \fmref{E-2-2}).}
\label{F-5}
\end{figure}

Finally we like to discuss our approximation for an
icosidodecahedron ($N=30$) of spins $s=5/2$ as for example
provided by \mofe\ \cite{MSS:ACIE99}. The total Hilbert space
has the dimension of $6^{30}\approx 10^{23}$, therefore exact
methods are nowadays (and in the foreseeable future) not
applicable. Figure~\xref{F-5} displays the approximate energy
levels in subspaces $S=0$ and $S=1$ for 8, 9, 10, and 11 used
rotational bands. Again the convergence is slow. Nevertheless,
using only a tiny fraction of all basis states -- $\sim 10^{5}$
out of $\sim 10^{23}$ -- for the diagonalization provides a much
richer spectrum than any of the methods used so far
\cite{GNZ:PRB06,CeZ:PTPS05,WAL:PRB07}. It is also interesting to
note, that the rotational band model seems to provide an
accurate estimate for the relative ground state energies in each
sector of total spin $S$ (crosses in \figref{F-5}), a fact that
has already been noted in \cite{ExS:PRB03}. Figure~\xref{F-5}
shows as well that the approximate spectrum is not yet
converged. This explains, why linear and even non-linear spin
wave theory for such frustrated systems yield poor results,
since these methods correspond to an approximate diagonalization
with two or a few more rotational bands.

\emph{Summary \& Outlook}---The approximate diagonalization in
terms of low-lying rotational band states provides a promising
description of the exact energy spectrum of mesoscopic
antiferromagnets. This method can be further improved by
employing point group symmetries, i.e. by diagonalizing in even
smaller and better adapted subspaces. To this end the ITO
technique has to be combined with point group symmetries, which
in general is rather demanding \cite{Wal:PRB00}.  Another route
to improvements is given by an approximate diagonalization in
terms of \emph{important} states as suggested by the authors of
Ref.~\cite{RoN:PRL07}. This improvement aims at more accurately
approximating some target states in order to obtain reliable
energy gaps at the costs of less accurate higher-lying excited
states.  Both improvements will be tested in future
investigations.

\emph{Acknowledgment}---We thankfully acknowledge computer time
at the Leibniz Computer Center Munich.


\end{document}